# QxEAI: Quantum-like evolutionary algorithm for automated probabilistic forecasting

Lizhi Xin[1], Kevin Xin[2*]

[1]Independent Researcher, Beijing, Beijing, People's Republic of China.

[2]Independent Researcher, Chicago, Illinois, United States of America.

*Corresponding author(s). E-mail: xin_kevin@outlook.com;
Contributing authors: lizhi_xin@yahoo.com;

**Abstract**

Forecasting, to estimate future events, is crucial for business and decision-making. This paper proposes QxEAI, a methodology that produces a probabilistic forecast that utilizes a quantum-like evolutionary algorithm based on training a quantum-like logic decision tree and a classical value tree on a small number of related time series. We demonstrate how the application of our quantum-like evolutionary algorithm to forecasting can overcome the challenges faced by classical and other machine learning approaches. By using three real-world datasets (Dow Jones Index, retail sales, gas consumption), we show how our methodology produces accurate forecasts while requiring little to none manual work.

## 1 Introduction

There are quite a few approaches on how to automate time series forecasting. There is the first systematic approach Box-Jenkins model, the more widely known auto-regressive (AR) models, exponential smoothing (ES), and auto ARIMA. [1-5] The Box-Jenkins model was outlined in the publication of *Time Series Analysis: Forecasting and Control* [2] by George Box and Gwilym Jenkins, which integrated the existing knowledge of the autoregressive and moving average methodologies. Their method relies on statistics, mostly on a linear relationship between the variables of what is being forecasted. [6-8] A wide range of AR methods have spurted out from the Box-Jenkins model, the widely used auto-regressive integrated moving average (ARIMA) method.

More recently the advances of machine learning and AI cannot be ignored, ML methods such as DL, autoML, and neural networks, which have shown more promising results than those of the traditional ARIMA methods. Machine learning methods for time series forecasting can fundamentally be categorized into three groups: linear modeling, deep learning, and automated machine learning. Linear modeling uses the native linear ways to forecast a time series; deep learning is a subordinate of neural networks that are based off of the biological brain and artificial neural networks, which are used to build a process to learn a large number of unknown inputs; and auto machine learning is solving a traditional machine learning task in a fully automatic or close to automatic way. [9]

Thus, that begs the question: Is it possible to automate time series forecasting by machine learning and AI? [10-13]

The challenges faced are: 1) the constantly changing external environment, 2) the internal uncertainty of events, and 3) the interaction between the external environment and an entity which triggers events to happen.

This work presents QxEAI, a methodology that produces probabilistic forecasts that utilizes a quantum-like evolutionary algorithm based on training a quantum-like logic decision tree and a classical value tree on a small number of related time series. Quantum-like superposed state represents all the possible solutions and the evolution algorithm then optimizes the most satisfactory solution to be taken.

The contributions of this paper are 1) we propose a quantum-like evolutionary algorithm (QxEAI) for probabilistic forecasting, which incorporates principles of evolution and principle of quantum superposition to tackle the environment and uncertainty, and 2) we show using three real-world datasets (Dow Jones Index, retail sales, and gasoline consumption), that this model is viable enough to produce probabilistic forecasts. In addition to providing more well-rounded accurate forecasts, our method has a few key advantages:

i. As the model learns through repetitive iterations, by evolution and quantum superposition, QxEAI formulates strategies that guide it to take certain actions with certain degree of belief, which are formulated in the form of logic decision trees that are optimized though evolution.

ii. Our approach does not assume uncertainty is "noise", something that we have to seek to eliminate or reduce. By learning with the mindset that uncertainty is inherent, we are able to provide forecasts for data that have little or no regularities, a case where many traditional forecasting methods don't perform as well.

The above points are what sets QxEAI apart from traditional forecasting approaches, our method enables optimal decision making under uncertainty by incorporating it alongside evolution and quantum-like beyond the forecast distribution. These probabilistic forecasts are of upmost importance in many real-world applications, contrary to traditional forecast methods which aim to make forecasts under most optimal and ideal conditions or presuming some trend.

By reeling together the principle of quantum superposition and evolution, QxEAI can find valuable information from raw data by pointing out a trend and probabilistic forecast. This entire process is mostly automatic, without any major need for manual intervention throughout the training and forecasting stages.

The structure of this paper as follows. Section 2 are the results. Section 3 details the QxEAI method. Section 4 discusses the architecture of the QxEAI model. Section 5 is the conclusion.

## 2 Results

We implemented our model from scratch using C#, and used a single Lenovo ThinkPad X1 Carbon to run all the experiments. Using this set-up, we trained and predicted on a small dataset of about 36 points, which can be completed in a few hours. QxEAI is set to a population of 300 individuals, evolving 80 generations, with the crossover probability at 70%, and the mutation probability at 5%.

### 2.1 Datasets

We use three datasets – Dow Jones Industrial Average, total US advance retail sales, and total gasoline consumption at US gas stations for our evaluations, all of them from May 2021 to April 2024 by monthly intervals. The first 24 months of data were used for training and the following 12 months of data for verification. Verification data is the "forecast" data that acts like it didn't happen yet but is "forecast" for verification purposes.

QxEAI utilizes the logic and value trees to produce more accurate probabilistic forecasts. Regarding forecasting for all three of the datasets applications results, QxEAI does two things: 1) for forecasting the one possibility that has the greatest chance of happening, and 2) for the trend curve. For forecasting future data points, QxEAI generates 10,000 possibilities and runs each one once, then it selects the one that incurs the most frequently. The one that is selected is then applied in the forecast process. For plotting the trend curve, QxEAI generates 1,000 possibilities also runs them once each, then averages them all, which is then the result of the trend curve.

### 2.2 Applications results

When there are certain regularities to be found, a certain function can be used to describe it, but when it comes to more volatile datasets or those engaged in a "random walk" some difficulties tend to arise. The traditional way has been to stubbornly attempt to find a certain function to describe these volatile datasets, but most times it has a hard time doing so just from raw data only. When certain regularities can't be obviously found, QxEAI has some slight advantages when it comes to tackling more challenging datasets.

Months of May 2021-April 2023 is training data, and the months of May 2023-April 2024 is the forecast data. The blue line is the original data points of each dataset, the yellow line in the train graph is the fitting curve while the yellow line in the forecast graph is the predicted curve, and the red line is the trend curve.

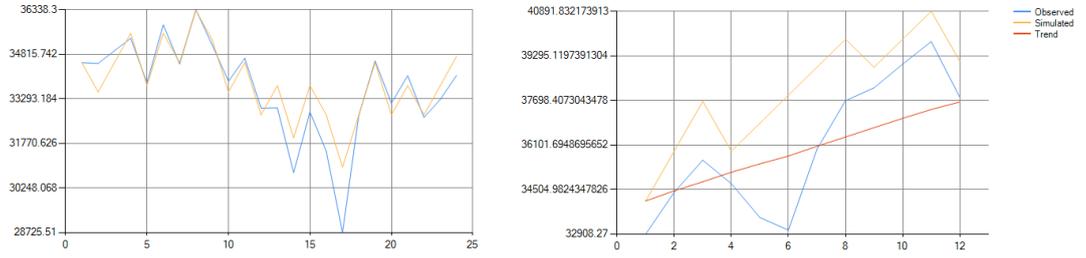

Figure 1. Dow Jones Index train on left and forecast on right.

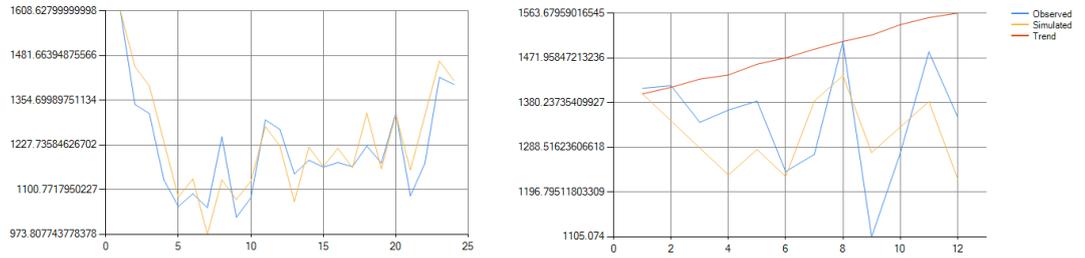

Figure 2. Retail sales train on left and forecast on right.

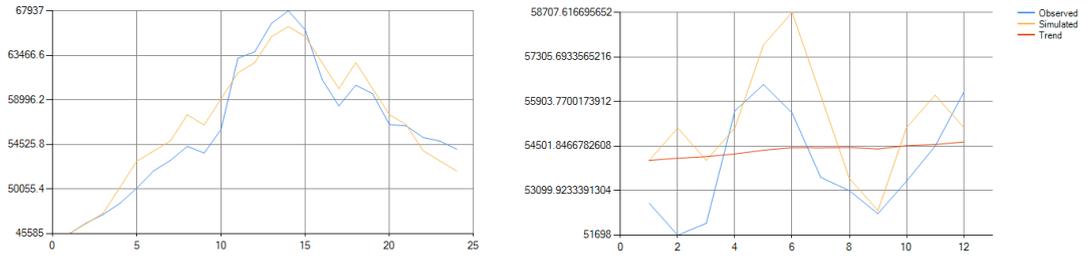

Figure 3. Gasoline consumption train on left and forecast on right.

Figure 1 (Dow Jones Industrial Index), Figure 2 (total US advance retail sales), and Figure 3 (total gasoline consumption at US gas stations) shows the training and forecast results by QxEAI. QxEAI fits good enough, the forecast data was almost spot on, and the trend curve was overall in the right direction. Table 1 shows the accuracy of the forecasts for all three datasets.

The MAPE and normalized RMSE metrics are defined as

$$\text{MAPE} = \frac{\sum_{i=1}^{N}|(o_i - p_i)/o_i|}{N} \times 100$$

$$\text{RMSE} = \frac{\sqrt{\sum_{i=1}^{N}(o_i - p_i)^2/N}}{\sum_{i=1}^{N}|o_i|/N}$$

Where $o_i$ is the observed value for item i, $p_i$ is the predicted value for item i, and N is number of intervals that are going to be predicted.

**Table 1** MAPE & RMSE Accuracy

|  | Dow Jones | Retail sales | Gasoline |
|---|---|---|---|
| MAPE | 6% | 6% | 3% |
| RMSE | 0.06 | 0.07 | 0.03 |

## 3 Methods

Time series can be described by consisting of its observed values and trend at a given time as (1-2).

$$\{(t_k, x_k, q_k)\}\ k = 1, \cdots, N \tag{1}$$

$$q_k = \begin{cases} 0, & x_k > x_{k-1} \\ 1, & x_k < x_{k-1} \end{cases} \quad (2)$$

Where $t_k$ is time, $x_k$ is the observed value, and $q_k$ denotes the trend. The trend is inherently unpredictable; thus, we superpose all the possible trend states (up or down) illustrated by the trend space as in (4) based on quantum superposition principle.

The quantum superposition principle is the principle that something can be in two states at the same time, the most famous example being Schrodinger's cat that is dead and alive simultaneously. [14] More precisely, all the possible states of a superposed state can potentially exist in a linear combination of its possible states before it is observed. Superposition's key idea is that the combination of multiple wave-like entities is not just the sum of all their effects individually. Quite the contrary, the waves interact in a way that leads to interference patterns, amplification, and even cancellation of some certain characteristics. [15]

The first postulate of quantum mechanics is "the state of an isolated physical system is represented, at a fixed time t, by a state vector ψ belonging to a Hilbert space $\mathcal{H}$ called the state space." [16] Thus, when something is in a superposed state, all the possible states can be expressed by a wave function, ψ, which can be expressed as a linear combination of the states of the observable as (3). [17]

$$\psi = c_1\psi_1 + c_2\psi_2 + \cdots + c_n\psi_n \quad (3)$$

There are corresponding observed values of $o_1, o_2, \ldots, o_n$, and once the measurement happens only one of these values $o_n$ can be observed with a certain degree of probability of $|c_n|^2$.

Then in order to "guess" the right trend as close as possible, we need to formulate certain strategies that'll guide us with actions (believe the trend will go up or believe it will go down) superposed as well, illustrated by the action space as in (5). The possible trend states and action states are all superposed in terms of Hilbert Space, and then the GP algorithm optimizes the best action to take based on the most maximized expected value.

For example, in the case of the stock market, the trend state is that the closing price will go up or go down, and the actions that can be taken are buy and sell. When traders trade in the stock market, all the traders' actions together determine the closing price of the stock, and in turn the uncertainty of the trend then affects the traders' actions, vice versa.

$$|\psi\rangle = c_1|q_1\rangle + c_2|q_2\rangle \quad (4)$$

Where $|q_1\rangle$ denotes the trend state going up, and $|q_2\rangle$ denotes the trend state going down. $\omega_1 = |c_1|^2$ is the objective frequency of the increase; $\omega_2 = |c_2|^2$ is the objective frequency of the decrease.

$$|\phi\rangle = \mu_1|a_1\rangle + \mu_2|a_2\rangle \quad (5)$$

Where $|a_1\rangle$ denotes the buy action, and $|a_2\rangle$ denotes the sell action. $p_1 = |\mu_1|^2$ is the subjective degrees of belief to buy; $p_2 = |\mu_2|^2$ is the subjective degrees of belief to sell.

Decision process: $\rho = |\phi\rangle\langle\phi| \xrightarrow{\text{decision}} \rho' = p_1|a_1\rangle\langle a_1| + p_2|a_2\rangle\langle a_2| \quad (6)$

Anytime someone makes a decision, their state of mind transforms from a pure state ρ into a mixed state ρ' as in (6), which is when they decide whether to do this or do that, in the case of trading on the stock market then it would be buy or sell, with certain degrees of belief. Essentially this decision-making state transformation is when one chooses to take an action from the pool of available ones with action $a_1$ being to buy with probability $p_1$ and action $a_2$ being sell with probability $p_2$.

This can be expressed in matrix form as in (7-8):

$$\rho = \begin{bmatrix} \rho_{11} & \rho_{12} \\ \rho_{21} & \rho_{22} \end{bmatrix} \xrightarrow{\text{diagonalization}} \begin{bmatrix} \lambda_1 & 0 \\ 0 & \lambda_2 \end{bmatrix} \xrightarrow{\text{normalization}} \rho' = \begin{bmatrix} p_1 & 0 \\ 0 & p_2 \end{bmatrix}$$
$$= p_1|a_1\rangle\langle a_1| + p_2|a_2\rangle\langle a_2| \quad (7)$$

$$|a_1\rangle = \begin{bmatrix} 1 \\ 0 \end{bmatrix}, |a_2\rangle = \begin{bmatrix} 0 \\ 1 \end{bmatrix}; \quad |a_1\rangle\langle a_1| = \begin{bmatrix} 1 & 0 \\ 0 & 0 \end{bmatrix}, |a_2\rangle\langle a_2| = \begin{bmatrix} 0 & 0 \\ 0 & 1 \end{bmatrix} \quad (8)$$

The pure state (quantum density matrix) ρ can be approximately constructed from eight basic quantum gates as (9).

$$\begin{cases} H = \frac{1}{\sqrt{2}}\begin{bmatrix} 1 & 1 \\ 1 & -1 \end{bmatrix} X = \begin{bmatrix} 0 & 1 \\ 1 & 0 \end{bmatrix} Y = \begin{bmatrix} 0 & -i \\ i & 0 \end{bmatrix} Z = \begin{bmatrix} 1 & 0 \\ 0 & -1 \end{bmatrix} \\ S = \begin{bmatrix} 1 & 0 \\ 0 & i \end{bmatrix} D = \begin{bmatrix} 0 & 1 \\ -1 & 0 \end{bmatrix} T = \begin{bmatrix} 1 & 0 \\ 0 & e^{i\pi/4} \end{bmatrix} I = \begin{bmatrix} 1 & 0 \\ 0 & 1 \end{bmatrix} \end{cases} \quad (9)$$

Although under most circumstances, no one is able to fully grasp whether the trend will go up or down, exactly the most challenging part to forecasting, so the only way is to "guess" as accurately as possible of whether the trend is going to be up or down. Just like as with the stock market, the trend of anything that is being recorded is inherently uncertain (random walk), and in turn the decisions the traders' make influences the trend state which causes the trend and the traders' actions to always be clouded with dual uncertainty. Regardless of how the trend ends up being, in order to select the best action to take, through GP it optimizes the best action available based on maximizing expected value to the greatest (searched ρ to which is the one that guessed "right" the most times).

Genetic programming (GP) is an algorithm based off evolutionary characteristics that uses random crossover, selection, and mutation to formulate an executable program that solves problems. By randomly generating a certain amount of a population, the GP algorithm learns the fitness of each individual and then by the principles of natural evolution for n number of generations to optimize a most "satisfied" solution. The fittest ones survive are the ones that are utilized, following the theory of natural evolution which states life has evolved through generations of selection, mutation, and crossover, the ones most adapted to the environment survive long enough to pass their genes off to the next generation. [18-22]

QxEAI optimizes the best actions by using value and logic trees. The value tree (10-12) is just a traditional function tree, and it calculates the difference of the absolute value of two points which will aid the logic tree.

(1) Operation set $F = \{+,-,\times,\div,\log,\exp,\sin,\cos\}$;

(2) Dataset $T = \{t, x_1, x_2, \dots, x_m\}$.

$$d_{t,t-1} = |x_t - x_{t-1}| \quad (10)$$

$$\text{valueTree} \xrightarrow{\text{compute}} d'_{t,t-1} \quad (11)$$

$$\text{fitness}_{\text{valueTree}} = -\sum_{k=1}^{n}(d'_{t,t-1} - d_{t,t-1})^2 \quad (12)$$

Where $d'_{t,t-1}$ is the absolute value of the difference of two points calculated by the value tree, and $d_{t,t-1}$ is the observed absolute value of the difference of two points of the market.

The logic tree is essentially a decision tree and with help from the value tree it guides which strategies to take with corresponding actions. The expected value under the current environment and the corresponding actions can be represented by a corresponding trend-action-decision table, shown in Table 2:

**Table 2** Trend-action-decision table

| Action | Trend $q_1$ | $q_2$ |
|---|---|---|
| $a_1$ | $p_1 \omega_1 * d_{t,t-1}$ | $-p_1\omega_2 * d_{t,t-1}$ |
| $a_2$ | $-p_2\omega_1 * d_{t,t-1}$ | $p_2\omega_2 * d_{t,t-1}$ |

Anytime a "decision" is made there is always an expected value. [23-25] The expected value of each individual one is the possible scenarios of what the outcome could be paired with the state of what is being observed, as in (13):

$$EV_t = \begin{cases} p_1\omega_1 d_{t,t-1}, \text{trend is up and QxEAI believes so with } p_1 \\ -p_2\omega_1 d_{t,t-1}, \text{trend is up and QxEAI doesn't so with } p_2 \\ -p_1\omega_2 d_{t,t-1}, \text{trend is down and QxEAI doesn't with } p_1 \\ p_2\omega_2 d_{t,t-1}, \text{trend is down and QxEAI believes with } p_2 \end{cases} \quad (13)$$

The fitness function of the logic tree is the sum of all the expected value of all the individual actions as (14):

(1) Operation set $F = \{+, *, //\}$;

(2) Dataset $T = \{H, X, Y, Z, S, D, T, I\}$.

$$\text{fitness}_{\text{logicTree}} = \sum_{t=1}^{n} EV_t \tag{14}$$

The purpose the fitness function is essentially an incentive system of reward and punishment. Before a decision is made and an action is chosen, there are four possibilities of outcome that exist. Of course, only one can happen, basically one out of the four scenarios of the expected value in Table 1. Therefore, if whatever that's being observed is trending upwards and the belief is that it's trending upwards, that's equivalent to a reward. But if something is trending downwards and the belief is that it's trending upwards, then a punishment is concurred. Vice versa with the other two scenarios. By learning historical data, the more rewards that are reaped then the more accurate chance we have of predicting the next trend. This also allows for no presumptions of the trend, the more times the right trend is "guessed" correctly the best strategies and actions are effectively evolved as a result. Generation after generation of evolution, the best strategy naturally arises, which is the ultimate goal of the fitness function. The best strategy that has evolved by natural selection is the one that can be utilized for future forecasting. [26-29]

## 4 Discussion

QxEAI applies quantum superposition to construct all the possible solutions, while GP optimizes the best one by the fitness function. There is a trend state (the closing price is up or down), and an action state (to buy or sell). Trend state and action state are in a state of dual uncertainty, for example, trend of the closing price is uncertain and the actions that can be taken are uncertain, all of which is constructed by quantum superposition in terms of Hilbert Space. QxEAI takes uncertainty into account and doesn't treat it as external noise; the ability to adapt to the real-world uncertainty allows it to fare much better than the classical ways.

As illustrated by Figures 1, 2, and 3 the QxEAI automatically learns from the historical data to find the trend and then use that trend for forecasting. QxEAI gives 2 results; one is the most likely to happen from 10000 runs as the forecast, and the other is the average of 1000 runs as the trend, and the results were fairly decent shown in Table 1.

## 5 Conclusion

We have shown that forecasting approaches based on evolution and principle of quantum superposition can drastically improve the accuracy of forecasting of the Dow Jones Index, retail sales, and gasoline consumption. Fundamentally, forecasting is attempting to predict how point A becomes point B: $A(t_n, x_n) \xrightarrow{m} B(t_{n+1}, x_{n+1})$, where m is all the different possibilities that could happen, essentially all the "paths" that could be taken. In reality only one "path" can actually be taken, but if somehow, we can know the complete information of which path will be taken by machine learning historical data, then a function will be produced and applied to forecast with certainty. If complete information is not obtained then the quantum-like way will give a probabilistic forecast with a trend curve and the curve that is most likely to happen. The quantum-like portion "deals with" the dual uncertainty of the external environment and the actions taken by the "decision makers", while the evolutionary portion "deals with" the interactions between the environment and the "decision makers", and by combining the two, QxEAI is able to produce forecasts under all types of scenarios.

In this paper we have shown an application to the stock market, retail sales, and gasoline consumption, with a method that requires little to none manual intervention. Thus, this answers the original questions that we started out with – the challenges of automating forecasting. Regardless of the dataset, it does so by loading of an Excel spreadsheet (small-sized datasets that contain less than 100 datapoints), without any prerequisite of programming and statistics knowledge for the end user. All that is required for the end user is domain knowledge of the dataset to be analyzed, and interestingly enough our QxEAI can be termed a "machine data scientist" by the functions it can perform.

## Declarations
**Funding -** This work received no funding.
**Conflicts of interest/Competing interests -** The authors declare no competing interests.
**Availability of data and material -** The data that support the findings of this study are available on request.
**Authors' contributions -** All authors conducted the research and contributed to the development of the model. L.X. contributed to the research from the aspects of machine learning, decision theory and wrote the code. K.X. wrote this manuscript and did data analysis. All authors reviewed the manuscript.

**Acknowledgements -** The authors would like to acknowledge retired Professor Houwen Xin for his expertise and prior guidance that he contributed to the writing of this paper.